\documentclass[10pt]{article}

\usepackage{amsmath}
\usepackage{amssymb}

\usepackage{graphicx}

\usepackage{cite}

\usepackage{color}

\usepackage[labelfont=bf,labelsep=period,justification=raggedright]{caption}

\makeatletter
\renewcommand{\@biblabel}[1]{\quad#1.}
\makeatother

\date{}

\begin{document}

\begin{flushleft}
{\Large
\textbf{Learning theories reveal loss of pancreatic electrical
  connectivity in diabetes as an adaptive response}
}
\\
Pranay Goel$^{1,\ast}$,
Anita Mehta$^{2}$
\\
\bf{1} Mathematics and Biology, Indian Insitute of Science Education and Research Pune, Pune, Maharashtra, India
\\
\bf{2} Department of Physics, S. N. Bose National Centre for Basic Sciences, Kolkata, West Bengal, India
\\
$\ast$ E-mail: pgoel@iiserpune.ac.in
\end{flushleft}

\section*{Abstract}
Cells of almost all solid tissues are connected with gap junctions
which permit the direct transfer of ions and small molecules, integral
to regulating coordinated function in the tissue. The pancreatic
islets of Langerhans are responsible for secreting the hormone insulin
in response to glucose stimulation.  Gap junctions are the only
electrical contacts between the beta-cells in the tissue of these
excitable islets. It is generally believed that they are responsible
for synchrony of the membrane voltage oscillations among beta-cells,
and thereby pulsatility of insulin secretion. Most attempts to
understand connectivity in islets are often interpreted, bottom-up, in
terms of measurements of gap junctional conductance. This does not,
however explain systematic changes, such as a diminished junctional
conductance in type 2 diabetes. We attempt to address this deficit via
the model presented here, which is a learning theory of gap junctional
adaptation derived with analogy to neural systems. Here, gap junctions
are modelled as bonds in a beta-cell network, that are altered
according to homeostatic rules of plasticity. Our analysis reveals
that it is nearly impossible to view gap junctions as homogeneous
across a tissue. A modified view that accommodates heterogeneity of
junction strengths in the islet can explain why, for example, a loss
of gap junction conductance in diabetes is necessary for an increase
in plasma insulin levels following hyperglycemia.

\section*{Introduction}
\begin{figure}[!ht]
\begin{center}
\centerline{\includegraphics[width=0.6\textwidth]{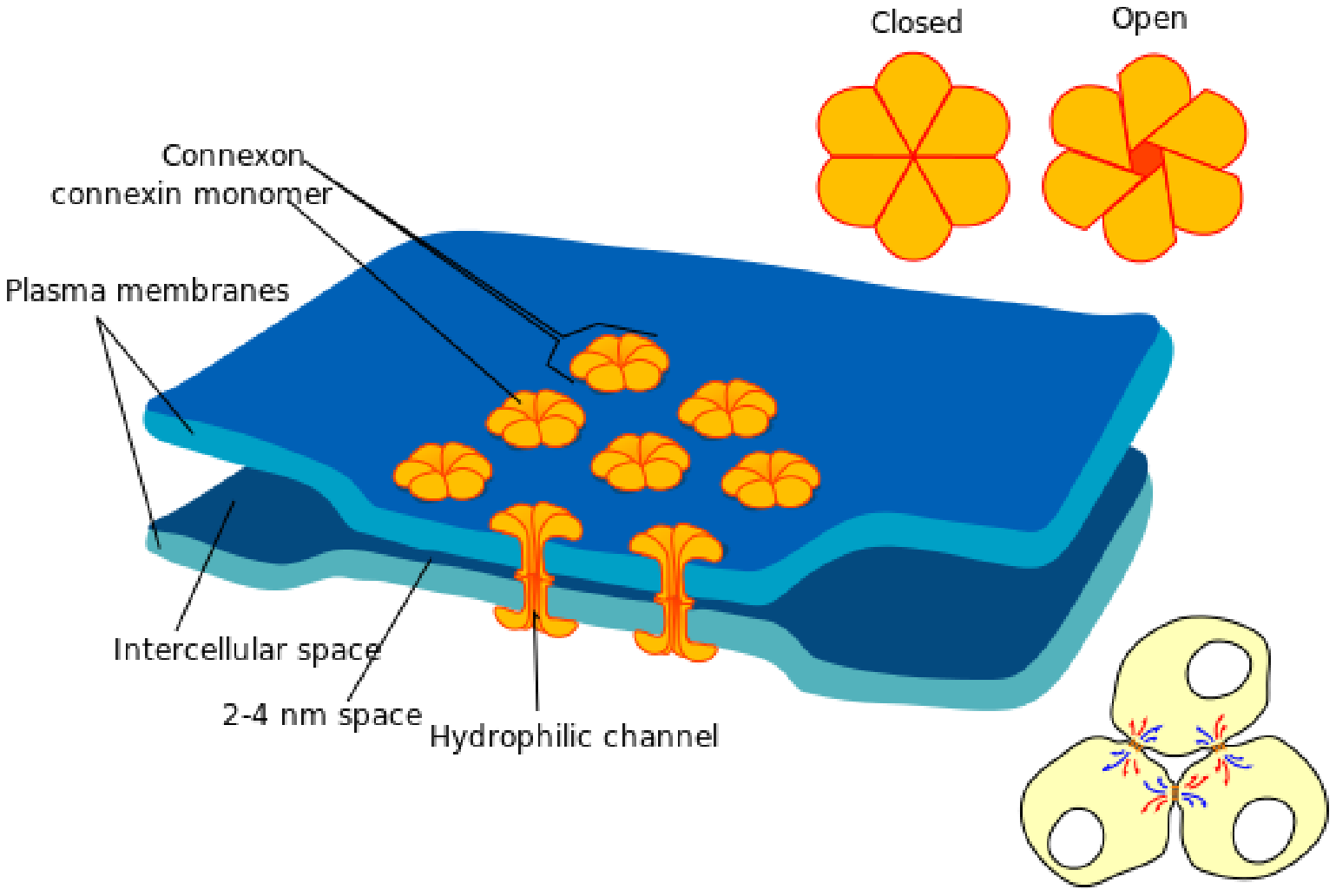}}
\caption{{\bf Gap junctions between cells permit intercellular
  communication.} Figure credit: Mariana Ruiz LadyofHats,
  http://en.wikipedia.org/wiki/File:Gap\_cell\_junction\_en.svg. }\label{gj}
\end{center}
\end{figure}

Gap junctions are clusters of intercellular channels between cells formed by the membrane proteins connexins (Cx), that mediate rapid intercellular communication via direct electric contact and diffusion of metabolites~\cite{goodpaul}. In excitable cells such as neurons, cardiac myocytes and smooth muscles, gap junctions provide efficient low-resistance pathways through which membrane voltage changes can be shared across the tissue.  Besides excitable cells, gap junctions are found between cells in almost every solid tissue~\cite{goodpaul}. Gap junctions are thus central to multicellular life~\cite{nicholson}, with numerous diseases linked to connexin disorders~\cite{willecke}, including type 2 diabetes mellitus~\cite{elke,ravier,medachatroom}. The islets of Langerhans in the pancreas are clusters of largely alpha-, beta- and delta-cells that respectively control secretion of the hormones glucagon, insulin and somatostatin central to energy regulation. Gap junctions form direct connections between beta-cells~\cite{orci73, boscomeda11,cabrera} in islets, and are important for normal glucose-stimulated insulin secretion (GSIS)~\cite{medachatroom,macd,head12}.  Gap junctions are generally believed to be important for coordinating the beta-cell electrical oscillations known as bursting, which in turn, can then support pulsatile insulin secretion~\cite{chansharing,macd,medachatroom}; this view is supported by theoretical studies~\cite{rinzel,shermansmolen,bss} as well. 
The conductance strength of gap junctions evolves by the insertion or deletion of connexin proteins (Fig.~\ref{gj}) into junctional plaques, and by altering the single-channel conductance and probability of channel opening~\cite{goodpaul}. Whether these molecular changes constitute a systematic adaptive response of the endocrine tissue to its metabolic environment remains to be investigated, in particular from a theoretical point of view.

As with many other excitable cells, the information content of bioelectric signals~\cite{levin} in islets is yet unclear. The mechanisms underlying bursting are well understood~\cite{glycophan,goel}; however, how those temporal properties regulate energy homeostasis is not. While slow (5 -- 15 minute period) bursts are generally thought to drive secretion at stimulatory concentrations of glucose, faster (periods less than 5 minutes) oscillations are also found, typically at sub-stimulatory (basal) glucose levels; the average calcium signal, however, is comparable in either case (such as in simulations from~\cite{glycophan}, not shown). The hypothesis that a synchronous bursting of beta-cells~\cite{medarojas,chansharing,macd} is essential to GSIS is guided by the observation of pulsatile insulin secretion from islets~\cite{bergsten} and in vivo~\cite{bergdiab}.  Gap junctions can certainly mediate synchrony in principle, as shown in both simulations~\cite{rinzel,chansharing} and experiments~\cite{calabrese,ravier}. Whether this is their role in vivo is debatable. In general, in vitro studies do not address this question completely, because they are typically carried out with glucose perifusion. Since glucose is microscopically delivered to beta-cells via a rich blood vessel supply in the islet in vivo, oscillator entrainment by junction coupling may be far less important than expected from experiments on isolated islets, especially if the beta-cells are not too heterogeneous in frequency~\cite{nunemaker}. In fact, Rocheleau et al.~\cite{piston} have performed experiments using a microfluidic chip taking care to see that glucose stimulates an islet only partially; they find partially propagated waves but not synchrony. Their result shows that gap junctions are limited in their ability to support uniform synchronization across the entire islet in the presence of a glucose gradient in the islet. It is possible that even with glucose micro-delivery as in vivo, synchronization may be a more local phenomenon than has been previously appreciated. Stozer et al.~\cite{stozer} have recently demonstrated that in islet slices only local synchronization is seen across groups of beta-cells. Another theory, different from one that anticipates gap junctions serve to synchronize an islet uniformly, thus appears to be necessary to explain some of the phenomena associated with insulin secretion, and it is this that we attempt in the rest of this paper.

A paradigm that is gaining increasing recognition is that   bioelectric and (epi-)genetic signaling are related as a cyclical dynamical system~\cite{levin}: membrane voltage activity induces changes in mRNA expression and transcriptional regulation, which in turn leads to altered membrane channel proteins. Here we develop a theory to study an adaptive response of gap junctions to islet firing activity. Bioelectric cues are encoded as bursting, these determine junctional conductance states, and junctions respond in turn by translation modifications that alter firing rates. In this way, electric and genetic components ``learn" from each other, iteratively. While learning is integral to neural systems and functionally beneficial at the level of a single individual, many studies have focused on the collective effects of [simple forms of] individual learning and decision-making, e.g. in populations of interacting individuals, or agents. Such distributed systems, exemplifying social or ecological group behavior, also share similarities with interacting systems of statistical physics, in the nature of the local ``rules" followed by the individual units as well as in the emergent behavior at the macro level. Game-theoretic approaches~\cite{game1,game2,game3} are sometimes brought to bear on such issues, their underlying idea being that the behavior of an individual (its ``strategy") is to a large extent determined by what the other individuals are doing. The strategic choices of an individual are thus guided by those of the others, through considerations of the relative ``payoffs" (returns) obtainable in interactive games. In this context, a stochastic model of strategic decision-making was introduced in~\cite{amjml99}, which captures the essence of the above-stated notion, i.e. selection from among a set of {\it competing} strategies based on a comparison of the {\it expected} payoffs from them. Depending upon which of the available strategic alternatives (that are being wielded by the other agents) is found to have the most favorable ``outcome" in the local vicinity, every individual appropriately revises its strategic choice.

Competition between prevalent strategies and adaptive changes at the individual level characterize the sociologically motivated model of~\cite{amjml99}. Given that these two features of competition and adaptation also generally occur across the framework of activity-induced synaptic plasticity, a translation of the notions in~\cite{amjml99} to the latter context was attempted in ~\cite{gmmam} and ~\cite{plosone}.  A model was delineated in ref.~\cite{gmmam} along these lines, with the types or weights of a plastic synapse taking the place of strategies. In the next subsections, we will extend these concepts to formulate a theory of `competing' gap junctions in a network.

\subsection*{Voltage gating of junctional conductance and homeostatic adaptation}

\begin{figure}[!ht]
  \centering
  \includegraphics[width=0.5\textwidth]{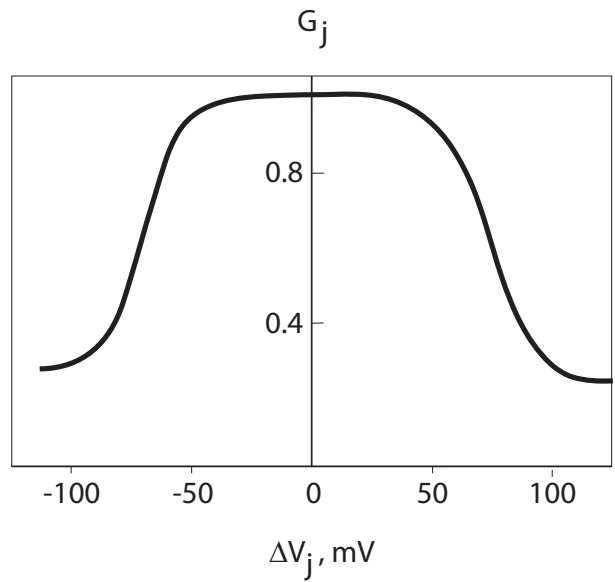}\\
  \caption{{\bf Voltage gating of Cx36 gap junctions, adapted from~\cite{teubner}.} Steady-state junctional currents from
    HeLa-Cx36 cell pairs indicate conductance, $G_j$, varies with
    transjunctional potential difference, $\Delta V_j$ . If two neighboring coupled cells fire nearly together, or do not simultaneously fire, trans-junctional conductance is high, but when one fires and the other does not conductance is low. This compensatory behavior inspires our {\em homeostatic} learning rule, see text.  }\label{vg}
\end{figure}

Gap junctions are known to adapt on at least two timescales:
trans-junctional currents are gated on a fast timescale of the order
of a few milliseconds to seconds in response to a trans-junctional
voltage difference ($\Delta V$)~\cite{goodpaul}. Voltage gated
currents of Cx36 channels (the connexin isoform relevant to
islets~\cite{ravier}), expressed in Xenopus oocytes and transfected
human HeLa cells, were recorded in~\cite{teubner}
(Fig.~\ref{vg}). Haefliger et al.~\cite{haefliger} have shown
hyperglycemia decreases Cx expression in adult rats. Paulauskas et
al.~\cite{bakauskas} have recently described a 16-state stochastic
model of gap junctional currents that are voltage gated by altering,
amongst other things, unitary single channel conductance and the
probability of opening~\cite{goodpaul,bakauskas4}. On much slower
timescales of hours to days, gap junctions are regulated by the events
that alter the insertion and deletion of channels in the junctional
plaque, connexin proteins synthesis, trafficking to the membrane and
degradation. We propose to study adaptation in gap junction strength
on slow timescales; this is the natural setting for a mean field
theory of gap junction modification, that is, over suitably long
periods that averages over cellular firing rates can be treated as
adiabatic.

Interestingly, the voltage-gated gap junction appears to conform to a homeostatic principle with respect to transjunctional current, $I_{gap}=g_{gap}\;\Delta V$: when $\Delta V$ is small, such as during synchronous bursting for example, gap junctional conductance is large, while a large $\Delta V$, as in anti-synchrony, is compensated with a small $g_{gap}$. That is, firing patterns $\Delta V$ result in changes in $g_{gap}$ that stabilize $I_{gap}$. We extrapolate from this argument to construct a {\em homeostatic learning rule} for (slow) modification of gap junctions, as described below.



\section*{Model and Results}

\subsection*{Model -- A learning theory of gap junctional adaptation}

\begin{figure}[!ht]
  \centering
  \includegraphics[width=0.7\textwidth]{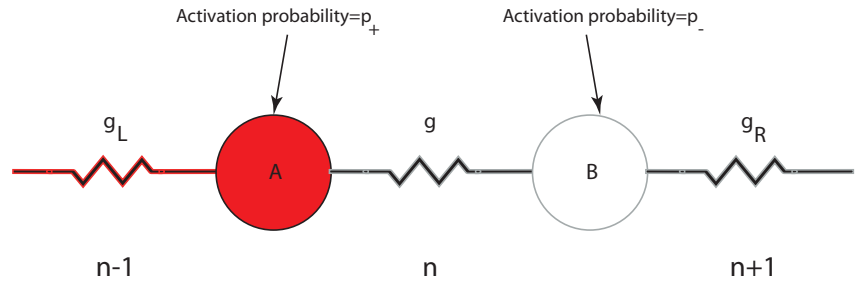}\\
  \caption{{\bf The bonds formalism of an islet.} $\beta$-cells, A and B,
    are dominated by gap junctions $g_L$ and $g_R$ respectively. Each junction ($g_L$ and $g_R$) can be in either strong (with probability $f_+$) or
    weak state (with probability $f_-$). A weak (strong) junction is likely to fire with a probability $p_-$ ($p_+$). The central
    gap junction $g$ is altered in response to the average potential
    difference of cells A and B, $\Delta V_{AB}$, across it, according to a specified learning rule, such as the homeostatic rule of Fig.~\ref{vg} that is considered here. For example, if cell A (red) here is assumed to fire in response to a strong $g_L$ (this occurs with probability $p_+$) while cell B is silent (the probability with which it could have been active is $p_-$) in response to a weak $g_R$, then the bond, g, will be weakened since $\Delta V_{AB}=1$.}\label{bonds}
\end{figure}

Our starting point is a model of competitive learning introduced in~\cite{amjml99} and applied, in~\cite{gmmam} and~\cite{plosone} to look at the optimisation of learning via a model of competing synapses. Proceeding by analogy, we consider a network consisting of $\beta$-cells connected by gap junctions, where the latter are treated as mutual neighbors if they are connected by a $\beta$-cell. In a one-dimensional formulation, each gap junction will thus be associated with two gap junctional neighbors. For simplicity the $\beta$-cells can be represented by binary threshold units, and the two states of the binary gap junction, which are inter-convertible by definition, are assumed to have different weights, which we label as `strong' and `weak' types.  A weak gap junction is characterized, for example, by fewer connexin proteins in the junctional plaque.   When the middle gap junction  is under consideration for a state update, the $\beta$-cells A and B (Fig.~\ref{bonds}) share this middle gap junction in common; thus, in comparing how often the two $\beta$-cells are found activated, one can factor out the influence of the common gap junction, when considering averages, and effectively treat the time-averaged activation frequency of either $\beta$-cell as being determined only by the single, {\it other} gap junction that the $\beta$-cell is connected to. This essentially implies that the state of $\beta$-cell A, say, can be considered quite reasonably as an ``outcome" to be associated with gap junction $n - 1$, and similarly with $\beta$-cell B and gap junction $n+1$; thus, $\beta$-cells can be thought of as taking on the identities of the respective gap junctions.

There are few general principles that can organize an argument to discuss plastic behavior in excitable cells; Hebb's postulate is one such. In common colloquialism this learning rule is stated as ``cells that fire together, wire together"; in other words, temporal association between pairs of firing neurons is successively encoded in synaptic coupling between those neurons. A Hebbian philosophy asserts that the direction of adaptation is such as to reinforce coordinated activity between cells. One can now set forth some rules governing the above weight changes, which may have a Hebbian or  anti-Hebbian flavor as the situation demands, and depend on the outcomes of the surrounding $\beta$-cells. Hebbian rules in the case of synaptic plasticity favour synchrony, so that e.g. a synapse is strengthened if its surrounding neurons fire or do not fire together; the opposite is the case with anti-Hebbian rules. In the present context, we use this concept analogously: for Hebbian rules, synchronous activity causes a strengthening of conductance while anti-synchronous activity causes a weakening of conductance.Thus, loosely speaking, two gap junctions adjacent to any given gap junction ``compete" to decide its type, and this continues to happen repeatedly across the entire network. Let us now consider the update dynamics of a single {\it effective} gap junction, that in some sense represents the average state of the whole network. To begin with, in such a picture, the outcomes are assumed to be uncorrelated at different locations, and treated as independent random variables, with the probability for activation being obtainable from the time-averaged activation frequency of the $\beta$-cell. Consistent with the situation described in the previous paragraph, that the effect of the common gap junction can be left out on average in comparing the outcomes of its connected $\beta$-cells, we associate, with each $\beta$-cell, a probability for activation at any instant that is {\it only} a function of the other neighboring gap junction, being equal to $p_+$ ($p_-$) for a strong (weak) type gap junction.

We now consider a mean-field version of the model. The idea behind the mean-field approximation is that we look at the average behavior in an infinite system. This, at one stroke, deals with two problems: first, there are no fluctuations associated with system size, and second, the approximation that we have made in ignoring the ``self-coupling" of the gap junction is better realized.
In the mean-field representation, every gap junction is assigned a probability (uniform over the lattice) to be either strong ($f_+$) or weak ($f_-$), so that spatial variation is ignored, as are fluctuations and correlations. This single effective degree of freedom allows for a description of the system in terms of its fixed point dynamics. The rate of change of the probability $f_+$, say, (which in the limit of large system size is equivalent to the fraction of strong units) with time, is computed by taking into account only the nearest-neighbor gap junctional interactions, via specific rules.

To design a transition rule for gap junctions that is consistent with a Hebbian theory, and  at the same time tunes gap junctional plasticity to voltage activity in the network, we mimic the homeostatic adaptation implicit in (fast) voltage-gating of conductance (Fig.~\ref{vg}): to reinforce synchronous activity conductance, changes must be directed towards a maximal state of conductance, while anti-synchronous activity is best served by a weakening of conductance. The {\em homeostatic learning rule} is summarised as follows: if $\beta$-cells (Fig.~\ref{bonds}) fire simultaneously $\Delta V_{AB}$ is zero and gap junction, $g$, strengthens to one, while if one $\beta$-cell fires but not the other, $\Delta V_{AB}$ is one and junction strength weakens to zero.

We write equations for the probability $f_+ (t+1)$ that the
intermediate gap junction (Figure~\ref{bonds}) is in the strong state, say, at time $t+1$
in terms of the same probability at time $t$, $f_+ (t)$, the
(complementary) probability that it was in the weak state at time $t$,
$f_- (t)$ and Prob($\Delta F$), the probability of a change in
strength of a given magnitude:
\begin{equation}
f_+ (t+1) = f_+ (t)\times \text{Prob}(\Delta F = 1) + f_-
(t)\times \text{Prob}(\Delta F = 1)
\end{equation}
The first term on the right hand side represents the probability that
the strong state at time $t$ stays strong at time $t+1$; since
 the gap junctions are binary, a strong junction cannot get any stronger.  Since $f_+ (t)+f_-
(t) = 1$, this reduces to the equation $f_S (t+1) = \text{Prob}(\Delta F =
1)$, independent of the initial state of the gap junction.

We now write down all possible scenarios for $\text{Prob}(\Delta F = 1)$: in words, these correspond to the sum of the following probabilities: (Prob that both $g_L$ and $g_R$ are in the strong state)$\times$(Prob that A and B both fire, AND both don't fire) + (Prob that both $g_L$ and $g_R$ are in the weak state)$\times$(Prob that A and B both fire, AND both don't fire) +
 (Prob that $g_L$ and $g_R$ are in disparate states)$\times$(Prob. that A and B both fire, AND both don't fire).

For example: if $g_L$ and $g_R$ (see Fig.~\ref{bonds}) are both strong -- with probability $f_+^2$ -- the firing pattern that leads to a strong middle junction, $g$, according to the homeostatic learning rule is when $\Delta V_{AB}=0$, i.e. either when both A and B fire simultaneously (probability, $p_+^2$), or both do not fire (probability, $(1-p_+)^2$). All such combinations are enumerated in Table~\ref{combotable}, this leads to an equation for the evolution of $g$:
\begin{eqnarray}
f_+(t+1) &= &f_+^2(t)\; (p_+^2 + (1-p_+)^2) \nonumber \\
&& +\; f^2_-(t) \;(p_-^2 +(1-p_-)^2) \label{eq:de}\\
&& +\; 2f_+(t)f_-(t)\;(p_+p_- + (1-p_+)(1-p_-) \nonumber
\end{eqnarray}

This evolution equation thus embodies that if $\beta$-cells (Fig.~\ref{bonds}) fire simultaneously, $\Delta V_{AB}$ is zero and gap junctions strengthen, while if $\Delta V_{AB}$ is one, junction strength weakens.

\subsection*{Results -- the steady state distribution of gap junctions}
The steady-state distribution of weak and strong junctions is obtained as the fixed point solution of Eq.~(\ref{eq:de}):
\begin{equation}
f_+^* = \frac{
-4p_-(p_+ - p_-) + 2(p_+ - p_-) + 1- \sqrt{-4(p_+^2-p_-^2)+4(p_+-p_-)+1}
}{4(p_+ - p_-)^2}.
\end{equation}
$f_+^*$ is stable in the entire $0<(p_-,p_+)<1$ domain. Perturbations
from $f_+^*$ relax at a rate $\lambda =1-\sqrt{1 - 4p_+^2 + 4p_+ - 4p_-  + 4p_-^2}$.


The physically reasonable condition on the firing probabilities is $p_- < p_+$. The minimum $f_+^*$ is $0.5$ which occurs for $p_- + p_+=1$ (Fig.~\ref{fig:pd2a}). That is, the theory predicts that in vivo at least half of the gap junctions in an islet will be of the strong type. It is possible for strong junctions to dominate the islet completely, $f_+^* \approx 1$, but this is seen to be an extreme scenario and requires either: $p_+$ is very low and $p_-$ as well, or $p_+$ is very high and $p_-$ is greater than about half. For the large part of the $(p_-,p_+)$ parameter space $f_+^*$ is predominantly between 0.5 and 0.7.

For low firing probabilities, such as for example $(p_-,p_+)=(0.1,0.15)$ the beta-cells A and B (Fig.~\ref{bonds}) seldom fire and $\Delta V_{AB}$ is invariably close to zero; $g$ therefore adapts towards the strong state. Likewise, when $p_-$ and $p_+$ are both high, such as for example at $(0.85,0.9)$  beta-cells A and B fire with a high rate and $\Delta V_{AB}$ is again close to zero and $g$ adapts towards the strong state. When the probabilities $p_-$ and $p_+$ are considerably different, however, for example when $(p_-,p_+)=(0.1,0.9)$ four possibilities arise: either A and B are both associated with weak (strong) junctions and $g$ adapts towards 1; or one of A or B is associated with a weak (strong) junction, but since one beta-cell then fires with a probability much larger than the other, $g$ adapts towards 0. Thus $f_+^*$ is close to half in this case ($g$ equally likely to be 0 or 1), as is the firing rate (Fig.~\ref{frate}).

We see thus that similar behaviour for the two gap junctions induces strengthening, while dissimilar  behaviour induces weakening, in line with the Hebbian viewpoint adopted above.

\section*{Discussion}

One major interest in developing a theory of gap junction adaption is to understand the changes in junctional conductance that take place in type 2 diabetes. It has been suspected from animal studies that loss of Cx36 is phenotypically similar to a prediabetic condition characterized by glucose intolerance, diminished insulin oscillations and first and second phases of insulin secretion, and a loss of beta-cell mass~\cite{bavamian07,hamelinmeda09,medachatroom,poto12}. Head et al.~\cite{head12} have recently confirmed this  {\em in vivo} via the observation that Cx36 conductance loss induces postprandial glucose intolerance in mice. These observations suggest that a loss of electrical connectivity in islets may underlie type 2 diabetes by disrupting insulin oscillations and reducing first-phase insulin secretion~\cite{head12,medachatroom}. Benninger et al.~\cite{benningerbasal} have found yet another effect that could be relevant to diabetes, that a loss of gap junctions in islets leads to increased basal (i.e. when  minimally stimulated by glucose) insulin release. If this were to hold in vivo it could explain hyperinsulinaemia as a result of gap junction loss as well, when steady state levels of circulating plasma insulin in diabetics continue to be high even in fasting conditions.

A word about dimensionalities  -- while we recognise that the geometries of real synaptic networks are complex and that they are embedded in three dimensions, our choice of working in one dimension is based as much on simplicity as on the absence of a reason to choose a more complex geometry. Working on a three-dimensional lattice would only increase the complexity of our algebra, while not really getting closer to the real geometry of synaptic networks, which are, as the name suggests, probably embedded on abstract graphs. However, the fact that we have worked in mean field (ignoring correlations and going to the limit of infinite systems) in a one-dimensional embedding makes our results less reliant on the embedding geometry than they otherwise might have been. We mean by this that while specific quantitative estimates might well be affected by the inclusion of more neighbours in higher dimensionalities, the qualitative outlines of our calculations will remain very similar. Our choice of mean field dynamics both in this case (as well as in the original learning model  of~\cite{gmmam})was very purposeful: in both cases, the exact geometries/connectivities of islets/synapses are imprecisely known, and infinitely variable. Under these conditions mean field theory is the tool most widely resorted to by modellers, since it is able to predict general features based on minimalistic assumptions.

\begin{figure}[!ht]
  \centering
  \includegraphics[width=0.7\textwidth]{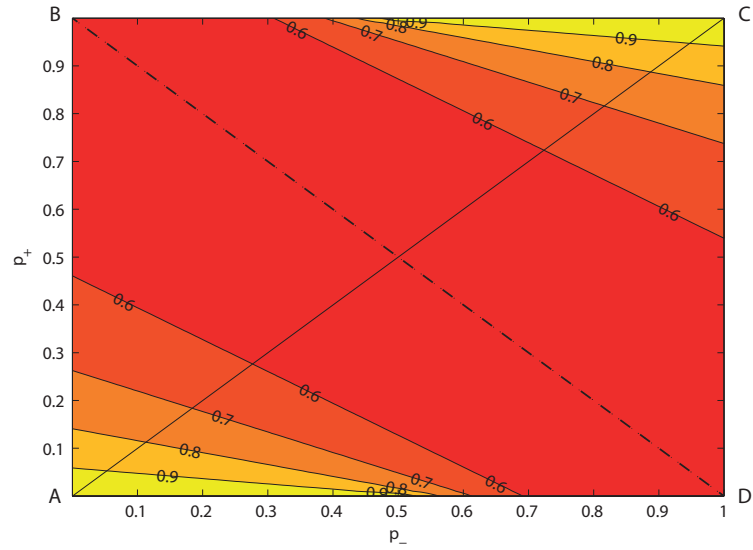}
  \caption{{\bf The $f_+^*$ contour plot in the $p_-$ -- $p_+$ plane.} The physically relevant ($p_-
    < p_+$) region is the triangle ABC above the line $p_+ = p_-$. $f_+^*=0$
    along BD, $p_+ +p_- =1$. The region near A where $f_+^*$ is close to $1$ represents healthy
    individuals while diabetics are assumed to lie along BD where $f_+^*=0.5$.}
  \label{fig:pd2a}
\end{figure}

\begin{figure}[!ht]
  \centering
  \includegraphics[width=0.7\textwidth]{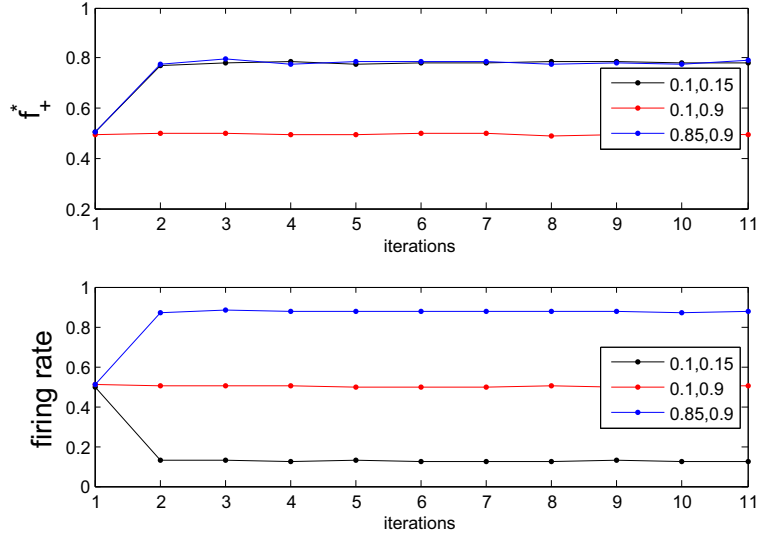}\\
  \caption{{\bf Evolution of gap junctions with network activity.} Beta-cells
    were initialized as firing (1) or not (0), and gap junctions as
    weak (0) or strong (1) with equal probability. 5000
    beta-cell--gap junction pairs (Fig.~\ref{bonds}) were iterated
    according to the learning rules described in the text. The legend
    indicates the ($p_-,p_+$) values for a computation. The top panel shows the evolution of the fraction of strong gap junctions, $f_+^*$, in the network. The bottom panel shows the corresponding fraction of beta-cells that are active. Note that $f_+^*$ as
    well as firing rate in the simulation are both 0.5 along $p_-+p_+=1$ as expected from the theory, Fig.~\ref{fig:pd2a}. A transition from health with low firing and high proportion of strong gap junctions (black curves) to diabetes takes place with degrading the gap junctions to increase firing rates (red curves).}\label{frate}
\end{figure}

The game-theoretic formalism presented here provides a high-level
explanation why a loss of junctional conductance would be necessary in
diabetes. In the healthy individual insulin secretion occurs
relatively sparingly, for a few hours at regularly spaced intervals
following glucose ingestion (breakfast, lunch and dinner). The low
firing rates in a healthy individual are accompanied by a high
proportion of strong gap junctions (that is, near the region marked by
A, Fig.~\ref{fig:pd2a}, where $f_+^*$ is close to 1). Diabetes is
associated with overnutrition among various other
factors~\cite{nathan}, and invariably involves combating an increased
glucose load~\cite{ada12,dcare12}. Several authors that proposed that
a substantial loss of Cx36 could ocur in type 2 diabetes (reviewed for
example in~\cite{hamelinmeda09}). Much of the evidence that connexins
expression or signaling are altered in models of type 2 diabetes comes
from rodents; however, because Cx36 is present in human islets, this
gives rise to the speculation (see e.g.~\cite{head12}) that a loss of
Cx36 gap junction conductance may occur in type 2 diabetes. Thus,
based on glucose intolerance measured in the conscious mouse Head et
al.~\cite{head12}, as well as
others~\cite{ravier,benzhang,speier,medachatroom}, have estimated that a
loss of nearly 50\% in junctional conductance could occur in
diabetes. In Fig.~\ref{fig:pd2a} the locus of a 50\% connectivity loss is
the line $p_+ + p_-=1$, where the fraction of strong gap junctions is
halved ($f_+^*=0.5$) but the firing rates are higher (Fig.~\ref{frate}). That is, the
islet stressed by an increase glycemic stimulation is forced to
respond with an increase in its firing and insulin secretion rate,
which it does by degrading strong gap junctions to weaker ones.

 In this way, the islet is able to accommodate a stimulus stronger
 than that for which its physiology had evolved. A change in $f_+^*$
 is accomplished largely through altering the probabilities of
 junction-induced firing, $p_+$ and $p_-$. As mentioned in the
 introduction, the classical view of diabetes is that it results from
 gap junction dysfunction. Instead, the game-theoretic theory we have
 presented relates a conductance decrease to an \textit {adaptive
   response of an islet that sacrifices strong gap junctions in order
   to maintain insulin control over hyperglycemia}.


 At the heart of our game-theoretic theory is its use of stochasticity
 in gap junction synchronisation. Classically, strong gap junctions
 entrain beta-cells to fire, the entire assembly is assumed to be
 fairly homogeneous in gap junction strength, and the resultant
 synchronous bursting is seen to be essential to GSIS. Our theory on
 the other hand, introduces the possibility that beta-cells coupled
 even to strong gap junctions may not fire, and likewise, weak gap
 junctions may induce simultaneous firing. Further, synchronous
 bursting, as well as the simultaneous \textit{absence} of bursting,
 induces stronger junctions, while antisynchrony weakens them. The
 result is that gap junctional strengths are constantly updated as a
 result of the synchronous or asynchronous bursting of beta-cells.  In
 other words, the core idea of our paper is that disparate firing
 patterns lead to changes in gap junctional strength -- which provides
 a hitherto unexplored scenario for synchrony. This then naturally
 leads to a situation where heterogeneity prevails in the distribution
 of gap junctional strengths in the islet. The heterogeneity of gap
 junctions in turn determines more complex patterns of activity in the
 network, beyond the simple categories of (anti-)synchronous bursting.

In principle it is possible to explain observations of junctional
strengths such as in~\cite{benningerbasal} individually, without
recourse to a general theory of gap junction function. Typically, a
lot of the focus is on studying the heterogeneity of beta-cells in an
islet. Indeed, Benninger et al. verify that different thresholds exist for
calcium excitations among the beta-cells of a (Cx36 null) islet, and
conclude therefore that beta-cells with high thresholds create
oscillator death~\cite{od} through gap junctions to decrease basal
secretion. The other question to ask, however, is: can heterogeneous
gap junctions within an islet shape the emergent properties of
bursting? Once the heterogeneity of the gap junctions themselves is
recognized as crucial, that leads, ipso facto, to an alternate view,
one in which changes in junctional conductance are seen as solutions
to an optimization problem. The essential ingredients of a theory of
gap junction adaptation include keeping track of the propensities with
which strong and weak junctions influence firing rates in beta-cells,
and transition rules that determine how gap junctions will respond to
local firing patterns. We have concentrated on learning rules that
embody homeostatic principles, which are a central feature of the
energy maintenance pathways of the body. However our general formalism
is certainly applicable to other forms of adaptation rules that may be
uncovered in future experiments.

We have constructed a theory that offers an alternative explanation to
the classical view that gap junctions primarily function to
synchronize beta-cells in an islet so the entire islet behaves like a
syncytium and a uniform period emerges. When gap junction adaptation
is considered, partial synchronization can occur even in networks fully
coupled with (strong) gap junctions. This learning framework predicts
in a natural fashion that a full synchrony across the islet is very
unlikely, that synchronization is a local phenomenon and happens
across a few groups of cells. Thus the view that emerges instead is that
the islet is sensitive to a glucose demand in secreting insulin and
uses gap junctions as a tuning parameter in this
adaptation. Paradoxically, an increase in secretion efficiency can
come not by strengthening junctions, but down-regulating them
instead. Thus, a lowered conductance need not necessarily be
interpreted as ``failing'' gap junctions. On the contrary, they are
judiciously adapting to the increased glucose load to cope with an
increased demand for insulin secretion.

At the moment there does not seem to be direct experimental evidence
that a reduction of gap junctions occurs in human type 2
diabetes. Additionally, although it is very attractive from a
theoretical viewpoint, it is not proven that gap junctions are altered
in response to altered islet firing activity in diabetes. Our model is
a complementary line of evidence, albeit theoretical, in these
directions. Further, the model makes another related prediction, that
gap junction expression and coupling strength are very likely to occur
as heterogeneous across the islet, in both health as well as
diabetes. If the naturally heterogeneous nature of gap junctions is
acknowledged, this could be critical in designing appropriate clinical
interventions, since connexins are potential targets for diabetes
therapy. Indeed, we hope that our work will be helpful to researchers
seeking to clarify the adaptive dynamics of gap junctions in diabetes.







\bibliography{gapbib}

\section*{Figure Legends}

\clearpage

\clearpage

\clearpage

\clearpage

\section*{Tables}
\begin{table}[!ht]
\caption{\bf{The probability of a gap junction adapting to a strong, high
  conductance state is determined by the current state of the bonds
  $g_L$ and $g_R$ (Fig.~\ref{bonds}).}}
  \begin{tabular}{ | c | c | c |}
    \hline
    $g_L$ & $g_R$ & $P$ \\ \hline
    Strong & Strong &  $f_+^2 \{p_+^2 + (1-p_+)^2\}$ \\ \hline
    Strong (Weak) & Weak (Strong) & $2f_+f_-\{p_+p_- + (1-p_+)(1-p_-)\}$ \\ \hline
    Weak & Weak &  $f^2_- \{p_-^2 +(1-p_-)^2\}$ \label{combotable}\\
    \hline
  \end{tabular}
\end{table}

\end{document}